\documentclass[conference, 10pt]{IEEEtran}
\usepackage[T1]{fontenc}
\usepackage[utf8]{inputenc}
\usepackage[english]{babel}
\usepackage{balance}
\usepackage{cite}
\usepackage{amsmath, mathtools}
\usepackage{amsfonts,amsthm,bm}
\usepackage{amssymb}
\usepackage{comment}
\usepackage{graphicx}
\usepackage{standalone}
\usepackage{dsfont}
\usepackage{mathtools}
\usepackage{color, colortbl}
\usepackage{soul}
\usepackage[normalem]{ulem}
\usepackage[acronym,shortcuts]{glossaries}
\usepackage{tikz}
\usepackage{arydshln}
\usepackage{bbm}
\usepackage{svg} 
\usepackage{algorithm,algorithmic}
\usepackage{balance}
\usepackage{adjustbox}

\usepackage{tikz}
\usepackage{pgfplots}
\pgfplotsset{compat=newest}
\pgfplotsset{plot coordinates/math parser=false}
\newlength\fheight
\newlength\fwidth
\usetikzlibrary{plotmarks,patterns,decorations.pathreplacing,backgrounds,calc,arrows,arrows.meta,spy,matrix}
\usepgfplotslibrary{patchplots,groupplots}
\usepackage{tikzscale}
\usetikzlibrary {patterns.meta}

\DeclareMathOperator*{\argmax}{\arg\!\max}

\newacronym{5g}{5G}{fifth-generation}
\newacronym{6g}{6G}{sixth-generation}
\newacronym{ai}{AI}{activation information}
\newacronym{aoi}{AoI}{age-of-information}
\newacronym{awgn}{AWGN}{additive white Gaussian noise}
\newacronym{bs}{BS}{base station}
\newacronym{dt}{DT}{data transmission}
\newacronym{fifo}{FIFO}{first in-first out}
\newacronym{fsa}{FSA}{framed slotted-ALOHA}
\newacronym{gf}{GF}{grant-free}
\newacronym{harq}{HARQ}{hybrid automatic repeat request}
\newacronym{iot}{IoT}{internet-of-things}
\newacronym{kpi}{KPI}{key performance indicator}
\newacronym{lsfc}{LSFC}{large-scale fading coefficient}
\newacronym{m2m}{M2M}{machine-to-machine}
\newacronym{map}{MAP}{maximum a posteriori probability}
\newacronym{mimo}{MIMO}{multiple input-multiple output}
\newacronym{minlp}{MINLP}{mixed integer non-linear programming}
\newacronym{ml}{ML}{maximum likelihood}
\newacronym{mmpc}{MMPC}{min-max pairwise correlation}
\newacronym{mmtc}{mMTC}{massive \ac{mtc}}
\newacronym{mtc}{MTC}{machine-type communications}
\newacronym{nr}{NR}{new radio}
\newacronym{noma}{NOMA}{non-orthogonal multiple access}
\newacronym{oma}{OMA}{orthogonal multiple access}
\newacronym{pb}{PB}{periodic beacon}
\newacronym{pdf}{PDF}{probability density function}
\newacronym{pia}{PIA}{partial information acquisition}
\newacronym{pima}{PIMA}{partial-information multiple access}
\newacronym{pomdp}{PO-MDP}{partially observable-Markov decision process}
\newacronym{ra}{RA}{random access}
\newacronym{rb}{RB}{reservation beacon}
\newacronym{rfid}{RFID}{radio-frequency identification}
\newacronym{sb}{SB}{scheduling beacon}
\newacronym{saloha}{SALOHA}{slotted ALOHA}
\newacronym{tdma}{TDMA}{time-division multiple-access}
\newacronym{urllc}{URLLC}{ultra-reliable low-latency communications}
\title{Partial-Information Multiple Access Protocol \\ for Orthogonal Transmissions}
\author{Alberto Rech\IEEEauthorrefmark{1}\IEEEauthorrefmark{3}, Stefano Tomasin\IEEEauthorrefmark{1}\IEEEauthorrefmark{2}, Lorenzo Vangelista\IEEEauthorrefmark{1}, and Cristina Costa\IEEEauthorrefmark{4} \\ 
\IEEEauthorrefmark{1}Department of Information Engineering, University of Padova, Italy.\\
\IEEEauthorrefmark{2}Department of Mathematics, University of Padova, Italy.\\
\IEEEauthorrefmark{3}Smart Networks and Services, Fondazione Bruno Kessler, Trento, Italy.\\
\IEEEauthorrefmark{4}S2N National Lab, CNIT, Genoa, Italy.\\ 
\small
\texttt{alberto.rech.2@phd.unipd.it, stefano.tomasin@unipd.it,} \\
\texttt{lorenzo.vangelista@unipd.it, cristina.costa@cnit.it}
}

\begin{document}
\maketitle

\begin{abstract}

With the stringent requirements introduced by the new \ac{6g} \ac{iot} use cases, traditional approaches to multiple access control have started to show their limitations. A new wave of \ac{gf} approaches have been therefore proposed as a viable alternative. However, a definitive solution is still to be accomplished. In our work, we propose a new semi-\ac{gf} coordinated \ac{ra}  protocol, denoted as \ac{pima},  to reduce packet loss and latency, particularly in the presence of sporadic activations. We consider a \ac{mtc} scenario,  wherein devices need to transmit data packets in the uplink to a \ac{bs}.  When using \ac{pima}, the \ac{bs} can acquire partial information on the instantaneous traffic conditions and, using compute-over-the-air techniques, estimate the number of devices with packets waiting for transmission in their queue. Based on this knowledge, the \ac{bs} assigns to each device a single slot for transmission. However, since each slot may still be assigned to multiple users, collisions may occur. Both the total number of allocated slots and the user assignments are optimized, based on the estimated number of active users, to reduce collisions and improve the efficiency of the multiple access scheme. To prove the  validity of our solution, we compare \ac{pima} to \ac{tdma} and \ac{saloha} schemes, the ideal solutions for \ac{oma} in the time domain in the case of low and high traffic conditions, respectively. We show that \ac{pima} is able not only to adapt to different traffic conditions and to provide fewer packet drops regardless of the intensity of packet generations, but also able to merge the advantages of both \ac{tdma} and \ac{saloha} schemes, thus providing performance improvements in terms of packet loss probability and latency.
\end{abstract}

\begin{picture}(0,0)(0,-520)
\put(0,0){
\put(0,0){\qquad \qquad \quad This paper has been submitted to IEEE for publication. Copyright may change without notice.}}
\end{picture}

\begin{IEEEkeywords}
\Acf{mtc}, \Acf{oma}, Partial-information, \Acf{iot}.
\end{IEEEkeywords}

\glsresetall

\IEEEpeerreviewmaketitle

\section{Introduction}\label{sec:introduction}

Two are the main categories of applications scenarios for \ac{mtc} foreseen to be enabled by \ac{5g} networks: \ac{urllc} and \ac{mmtc}. Both of them show several distinct features and challenges, that stem from the necessity to address the needs of the emerging \ac{iot} applications and services.  The performance requirements of each category are quite challenging, e.g., \ac{urllc} use cases target a maximum latency of 1~ms and reliability of 99.99999\%  (e.g. mission-critical applications), while \ac{mmtc} scenarios require supporting devices with a density up to 1 million devices per square km (e.g. Industrial IoT).  
For such demanding targets, that will eventually be more strict in \ac{6g} networks, several technical advances should be adopted at all layers. In particular, designing appropriate multiple access techniques and protocols is particularly relevant due to their impact on the limited resources and capabilities of the transmitting devices. 
Before the advent of \ac{5g}, adopted multiple access protocols were based on resource requests and grants \cite{Centenaro17} thus incurring on signaling overhead. \Ac{gf} approaches address this issue by allowing users to transmit the data immediately, without the need to wait for the grant approval of the \ac{bs}. \ac{gf} approaches include several different techniques, which can be classified into uncoordinated and coordinated \ac{ra} \cite{ElTanab21}. 

\paragraph*{Uncoordinated \ac{ra}}  these approaches can deal effectively with collisions while requiring limited communication overhead. Users transmit at random time instants, and specific techniques are adopted at the receiver to mitigate the effects of collisions. Among the uncoordinated \ac{ra} solutions, \ac{noma}  \cite{Saito13} has been widely advocated as the most promising and as an alternative to grant-based \ac{oma}.
However, \ac{noma} requires advanced pairing and power allocation techniques, as well as powerful channel coding and interference cancellation mechanisms that only partially mitigate the collision effects. 
Under these conditions, the \ac{bs} may become prohibitively complex to serve a large number of users.
In recent years, unsourced  \ac{ra} has been proposed as an effective solution to manage a massive number of devices \cite{Polyanskiy17}. In this paradigm, at any time, a fraction of devices transmit simultaneously, making use of the same channel codebook. The receiver decodes arriving messages without knowing the identities of the transmitters. Although this approach is very effective in managing many users, good performance can be achieved only for very small payloads and with high-complexity massive \ac{mimo} receivers \cite{fengler2021non, decurninge2020tensor}.

\paragraph*{Coordinated \ac{ra}} these solutions typically divide time into slots, each with the duration of one packet.  \Ac{saloha} is the simplest and most widely adopted coordinated \ac{ra} protocol: users transmit at the beginning of the first slot available after packet generation. When collisions occur, a random delay is added before re-transmitting the collided packets. Typically, the random delay has the same statistics for all users, and the coordination is limited to slot synchronization.
One of its variants, the \ac{fsa} protocol, has been widely adopted in \ac{rfid} systems \cite{Lee05, Su16}. In \ac{fsa}, time is divided into frames, and each of these is split into slots. Each user is allowed to transmit in only one slot per frame. Whenever an uplink packet is generated, the user postpones its transmission until the next frame, then it selects a specific slot uniformly at random. Unlike the standard \ac{saloha}, this protocol reduces collisions. Coordinated \ac{ra} solutions are particularly useful when user activations are highly correlated, for example as a result of correlated underlying traffic generation \cite{3GPP37868}.
Also, re-transmissions (with the consequent accumulation of packets in user queues) may yield correlated transmissions among different users. If on the one hand, such correlation further increases the chances of collisions; on the other hand, it can be exploited to indirectly coordinate \ac{ra}. In the literature, correlation-based schedulers have recently gained attention as a possible breakthrough for multiple access in \ac{mtc}. Such schemes typically rely on the knowledge of traffic generation statistics \cite{Kalor18, Moretto21a}, or learn the traffic correlation by exploiting the capabilities of machine learning tools \cite{Rech21, Sebab20}.
Lastly, an extreme case of coordinated \ac{ra} is \textit{fast uplink grant} \cite{Ali19}, wherein the \ac{bs} schedules one slot for each user, without any resource request. Note that in this case, the access randomness is removed, while coordination remains. Slots are usually shared by multiple users, thus collisions may still occur in case of simultaneous transmissions.

In this paper, we introduce a new semi-\ac{gf} coordinated \ac{ra} protocol, named \ac{pima}. In \ac{pima}, time is organized into frames of {\em variable length}, each divided into two sub-frames. The first is the \ac{pia} subframe, where active users (having packets to transmit) send a signal to the \ac{bs}. Using a compute-over-the-air approach \cite{Goldenbaum13,Liu20,Zhu21}, the \ac{bs} measures the received power and estimates the number of active users. Based on this knowledge, the \ac{bs} then assigns one slot to each user for the transmission in the \ac{dt} sub-frame. 
We stress that with respect to other two-step \ac{ra}-access schemes consisting of preamble and data transmission stages \cite{Centenaro17}, \ac{pima} acquires only partial information on the activation statistics in the \ac{pia} sub-frame, avoiding to reveal the users' identities.  

The rest of the paper is organized as follows. In Section~\ref{sec:systemmodel}, we first introduce the system model and the packet generation processes. Then, in Section~\ref{sec:protocol}, we describe the frame structure and the \ac{pima} protocol. Section~\ref{sec:numest} provides the procedure used to perform the partial information estimation, while  Section~\ref{sec:frameeff} presents the scheduling optimization problem conditioned on this information. In Section~\ref{sec:numericalresults} we discuss the numerical results and compare \ac{pima} with conventional \ac{tdma} and \ac{saloha} schedulers. Finally, in Section~\ref{sec:conclusions} we draw some conclusions.
\paragraph*{Notation} Scalars are denoted by italic letters, vectors, and matrices by boldface lowercase and uppercase letters, respectively. Sets are denoted by calligraphic uppercase letters and $|\mathcal{A}|$ denotes the cardinality of the set $\mathcal{A}$. $\mathbb{P}(\cdot)$ denotes the probability operator and $\mathbb{E}[\cdot]$ denotes the statistical expectation.

\section{System Model}\label{sec:systemmodel}

\begin{figure*}
	\centering
	\includegraphics[width = 0.7\textwidth]{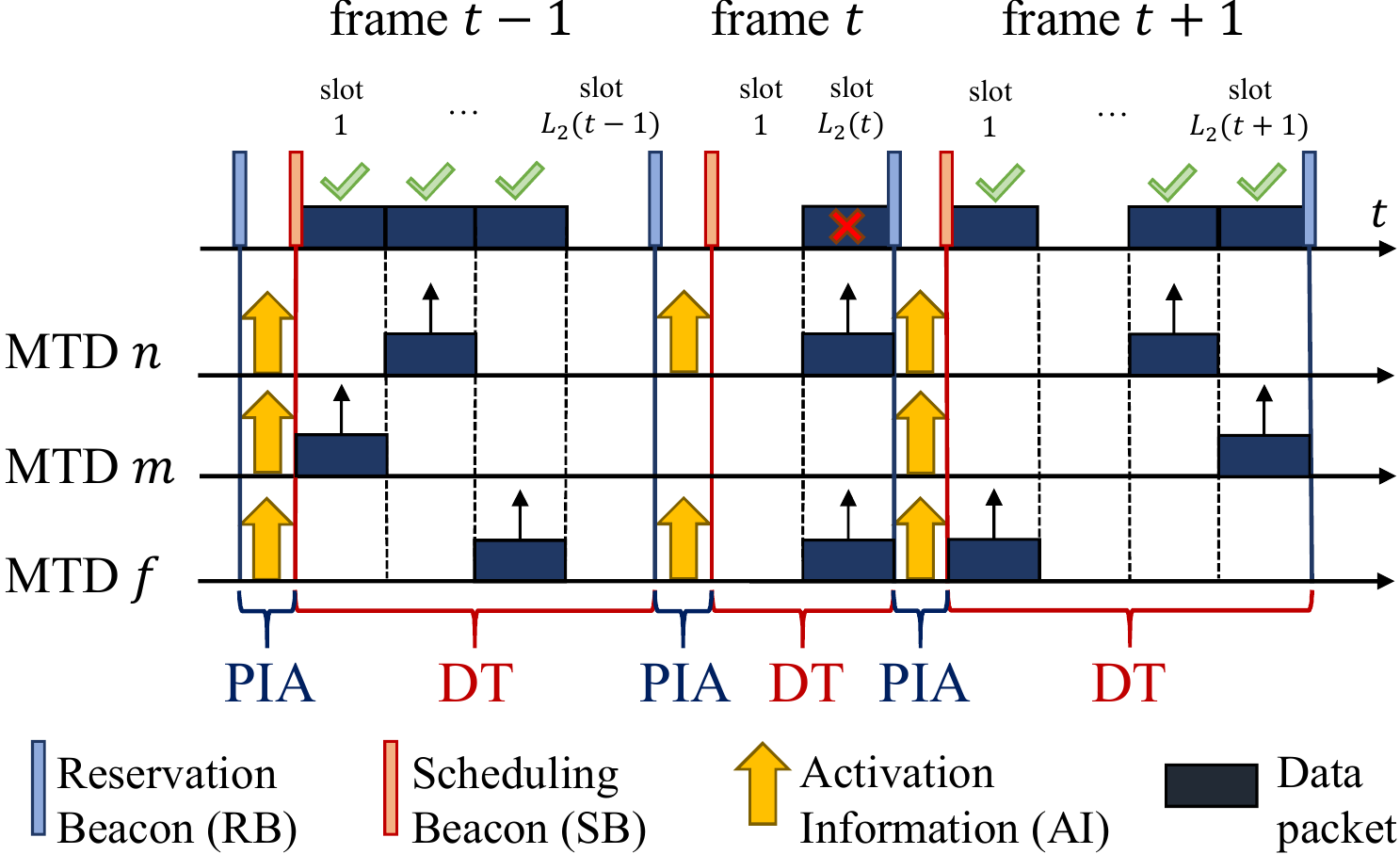}
	\caption{Example of the \ac{pima} protocol and its frame structure.}
	\label{fig:framestructure}
\end{figure*}

We consider an uplink multiple access scenario with $K$ users transmitting in the uplink to a common \ac{bs}. We assume that the value of $K$ is known at the \ac{bs}, while this knowledge is not needed by any user.

Time is divided into {\em frames}, each comprising an integer number of {\em slots} and an additional short time interval, whose purpose is described in the following. Each slot has a fixed duration, while each frame comprises a different number of slots. Perfect time synchronization at the \ac{bs} is assumed, thus each user can transmit signals with specific times of arrival at the \ac{bs}.
Users transmit packets, each with the duration of a slot, and each user can transmit at most one packet per frame.  In the following analysis, $\tau$ denotes the generic time instant, $t$ the frame index, and $\tau_0(t)$ the start time of frame $t$. In the considered setup, the same slot is in general assigned to multiple users for transmission.

\paragraph*{Channel}
Due to the scheduling of the same slot to multiple users, collisions between packets may occur. We assume that when two or more users transmit in the same slot, a collision occurs, preventing the decoding of all collided packets by the \ac{bs}.
Successful transmissions are acknowledged by the \ac{bs} at the beginning of the following frame, and in the event of collisions, users retransmit their packets in the following frame. We assume that the \ac{bs} always correctly decodes the received packets in slots without collision, thus the channel does not introduce other sources of communication errors.

\paragraph*{Packet Generation and Buffering}
Packets generated in frame $t$ by user $k$ are stored in its buffer and transmitted in the next slots, according to a \ac{fifo} policy.
We assume that all users have a limited buffer capacity $\bar{B}$.
To ensure data freshness, whenever a new packet is generated while the buffer is at full capacity, the oldest packet is dropped.
Let $B_k(\tau)$ be the length of user $k$ queue at time $\tau$. We also define the \textit{activation vector} as $\bm{B}(\tau) = [B_1(\tau), B_2(\tau), \ldots, B_k(\tau)]$. If $B_k(\tau) > 0$, the buffer of user $k$ is non-empty at time $\tau$, and $k$ is said to be \textit{active}, instead, if $B_k(\tau)= 0$, its queue is empty and user $k$ is considered \textit{inactive}. The total number of active users at the beginning of frame $t$ is $\nu(t)$.
Finally, we define the \textit{activation probability} of user $k$ in frame $t$ as 
\begin{equation}
\phi_n(t) = \mathbb{P}(B_k(\tau_0(t))>0).
\end{equation}

\section{Partial-Information Multiple Access}\label{sec:protocol}
In this section, we provide a detailed description of the proposed \ac{pima} protocol.
Each frame is divided into two {\em sub-frames}, namely the {\em \ac{pia} sub-frame} and the {\em \ac{dt} sub-frame}. The \ac{pia} sub-frame is used to estimate (at the \ac{bs}) the number of currently active users. Based on this information, the \ac{bs} decides the duration (in slots) of the \ac{dt} sub-frame and assigns each user to one slot, for possible uplink data transmission. Such scheduling information is transmitted in unicast to each of the users at the end of the \ac{pia} sub-frame. An example of \ac{pima} is shown in Fig.\ref{fig:framestructure}.

\subsection{Partial Information Acquisition Sub-Frame}

The beginning of frame $t$, and thus of the \ac{pia} sub-frame, is triggered by the \ac{rb}, which is transmitted in broadcast by the \ac{bs} to all users. \acp{rb} are transmitted to mark the start of the \ac{pia} sub-frame and contain the acknowledgments of the transmissions that occurred during the previous frame. Moreover, the \ac{rb} allows each user to estimate the \ac{lsfc} coefficient of the channel to the \ac{bs}, denoted as $g_k$. 
In the \ac{pia} sub-frame the \ac{bs} obtains the estimate  $\hat{\nu}(t)$ of  the number of active users $\nu(t)$. While this problem has already been discussed in the literature and solved with deep neural networks \cite{khan2022enumeration}, here we propose a novel low-complexity estimate inspired by computing over-the-air \cite{Goldenbaum13}.
For a duration $L_1$, the users transmit signals to let the \ac{bs} estimate the number of active users $\nu(t)$. More details on this procedure are provided in Section~\ref{sec:numest}.

At the end of the \ac{pia} sub-frame, the \ac{bs}, knowing $\hat{\nu}(t)$, schedules the transmissions for the next sub-frame.
Let $\bm{q}(t) = [q_1(t), \ldots, q_K(t)]$ be the \textit{slot selection vector}, collecting the slot indices assigned to each user; then, the length $L_2(t)$ of the \ac{dt} sub-frame can be derived from $\bm{q}(t)$ as
\begin{equation}
L_2(t) = \max_{\substack{k}} q_k(t).
\end{equation}
To end the \ac{pia} sub-frame and trigger the beginning of the following \ac{dt} sub-frame, the \ac{bs} transmits the \acp{sb}, which contains the slot selection vector $\bm{q}(t)$~\footnote{To maintain synchronization, inactive users could a) wake up and wait for the next downlink \ac{rb} when generating a packet, or b) always wake up when \acp{rb} and \acp{sb} are transmitted (this can be achieved by collecting the timing information in the beacons).}.

\subsection{Data Transmission Sub-Frame} 

In the \ac{dt} sub-frame, users transmit their packets, according to the scheduling set by the \ac{bs} in the \acp{sb}. 
 
If a packet is generated by the user $k$ during the \ac{dt} sub-frame, the packet is delayed and transmitted in the following frame. This feature is needed to ensure low collision probability, as the \ac{dt} frame length is derived only based on the number of users active in the \ac{pia} sub-frame.

\section{Estimation of The Number of Active users}\label{sec:numest}

To obtain an estimate of the number of active users at the \ac{bs}, each active user transmits an \ac{ai} signal of duration $L_1$ immediately after receiving the \ac{rb}. The  transmit power is such that the signal from each user is received with the same (unitary) power at the \ac{bs}. Note that we are neglecting here the propagation time between the \ac{bs} and the user, which can be easily accommodated by considering a transition (silent) time between the \ac{rb} and \ac{ai} transmissions.  

In particular, given a total system bandwidth $W$, we assume that each user transmits $M_1 = WL_1$ complex Gaussian symbols in the \ac{pia} sub-frame with zero mean and power $1/g_k$. 
The \ac{bs} then measures the total received power and estimates the number of active users. The set of users transmitting the \ac{ai} signals during the \ac{pia} sub-frame is 
\begin{equation}
\mathcal{K}_{\rm a}(t) = \{k : B_k(\tau_0(t))>0\},
\end{equation}
with $|\mathcal{K}_{\rm a}(t)| = \nu(t)$. The \ac{bs} does not know the identity of the active users, since the \ac{ai} signals do not contain such information, to make them shorter.

Letting the received samples at frame $t$ be $\tilde{\gamma}_\ell(t)$, $\ell=1, \ldots, M_1$, the estimated total power is
\begin{equation}\label{defPt}
\hat{P}(t) =  \frac{1}{M_1} \sum_{\ell=1}^{M_1} \tilde{\gamma}_\ell(t) =  \frac{1}{M_1} \sum_{\ell=1}^{M_1}\left| w_\ell(t) + \sum_{k \in \mathcal{K}_{\rm a}(t)} \gamma_{k,\ell}(t)   \right|^2, 
\end{equation}
where $\gamma_{k,\ell}(t)$ is the signal received from  user $k$ and $w_\ell(t)$ is the \ac{awgn} term with zero mean and variance $\sigma_w^2$. 

Let us indicate the \ac{pdf} of the received power given that $\nu(t)=b $ users are active as $p_{\hat{P}|\nu(t)}(a|b)$, and the probability that $\nu(t)=b$ users are active as $p_{\nu(t)}(b)$. The \ac{map} estimate of the number of active users is then
\begin{equation}\label{hnu}
\hat{\nu}(t) = \argmax_{b}\; p_{\hat{P}|\nu}(\hat{P}(t)|b)\;p_{\nu}(b).
\end{equation}

The value of $\hat{\nu}(t)$ is obtained from \eqref{hnu} by  splitting the set of real numbers into $K$ properly designed intervals ${\mathcal I}(b) = [\epsilon_{b-1}, \epsilon_{b}]$, $b=1, \ldots, K$, and finding the region where  $\hat{P}(t)$ is falling. Note that the first and last intervals are special, as for the first we have $\mathcal I(0) = [0, \epsilon_{0}]$, while for the last we have  $\mathcal I(K) = [\epsilon_k, \infty]$.  Decision regions may have different lengths, thus providing different estimations according to $b$. 

Optimal decision regions $\mathcal{I}(b)$, $b = 1, \ldots, K$ are intervals with boundaries at the intersections between the adjacent Gaussian curves; in particular, the boundary $\epsilon_b$ between interval $\mathcal I(b)$ and $\mathcal I(b+1)$ must solve
\begin{equation}\label{decreg}
    \frac{p_{\nu}(b)}{\sigma_{P}(b)\sqrt{2\pi}} e^{-\frac{(\epsilon_b-\bar{P}(b))^2}{\sigma^2_{P}(b)}} = \frac{p_{\nu}(b+1)}{\sigma_{P}(b+1)\sqrt{2\pi}} e^{-\frac{(\epsilon_b-\bar{P}(b+1))^2}{\sigma^2_{P}(b+1)}}.
\end{equation}
Equation \eqref{decreg} is equivalent to the quadratic equation $A\epsilon_b^2+ B\epsilon_b+ C = 0$, with 
\begin{equation}
\begin{split}
    A &= \frac{1}{\sigma^2_{P}(b+1)} - \frac{1}{\sigma^2_{P}(b)},\\
    B &= \frac{2\bar{P}(b)}{\sigma^2_{P}(b)} - \frac{2\bar{P}(b+1)}{\sigma^2_{P}(b+1)},\\
    C &= \frac{\bar{P}(b+1)^2}{\sigma^2_{P}(b+1)} - \frac{\bar{P}(b)^2}{\sigma^2_{P}(b)} - \log\left(\frac{p_{\nu}(b+1)\sigma_{P}(b)}{p_{\nu}(b)\sigma_{P}(b+1)}\right).
\end{split}
\end{equation}
Among the solutions of the quadratic equation, we must choose that falling between $\bar{P}(b)$ and $\bar{P}(b+1)$.

\paragraph*{Estimation Error Probability}
We define the average error probability
\begin{equation}\label{pe}
\bar{p}_{\rm e} = \mathbb{E}[\nu \neq \hat{\nu}] = \sum_b p_{\rm e}(b) p_{\nu}(b),
\end{equation}
where $p_{\rm e}(b) = \mathbb{P}[\hat{\nu}\neq b| \nu = b]$.

Assuming $\gamma_{k,\ell}(t)$ and $w_\ell(t)$ \textit{independent identically distributed (i.i.d.)} $\forall t, k, \ell$, as the square modulus of the sum of complex Gaussian random variables is exponentially distributed, $\hat{P}(t)$ follows an Erlang distribution with shape $M_1$ and rate $1/(\nu(t) + \sigma_w^2)$.
For large values of $M_1$, $\hat{P}(t)$ can be well approximated as a Gaussian variable with mean $\bar{P}(\nu(t))=\nu(t) + \sigma_w^2$ and variance
\begin{equation}
    \sigma_P^2(\nu(t)) = \frac{[\nu(t) + \sigma_w^2]^2}{M_1}.
\end{equation} 
Then, the conditional estimation error probability is
\begin{equation}\label{eq:perr}
\begin{split}
p_{\rm e}(b) &= \mathbb P[\hat{P}(t) \notin {\mathcal I}(b)| \nu(t) = b]\\
&={\rm Q}\left(\sqrt{M_1}\frac{\epsilon_b-b+\sigma_w^2}{(b+\sigma_w^2)}\right)\\
&\quad\quad+{\rm Q}\left(\sqrt{M_1}\frac{-(\epsilon_{b-1}-b+\sigma_w^2)}{(b+\sigma_w^2)}\right),
\end{split}
\end{equation}
where ${\rm Q}(\cdot)$ is the tail distribution function of the standard normal distribution. For $b=0$ and $b=K$, the first and second terms in \eqref{eq:perr} are zero, respectively. 

To minimize $L_1$, we should derive the minimum $M_1$ that guarantees the achievement of a target $p_{\rm e}$. However, the computation of the optimal decision regions from \eqref{pe} is in general quite complex, due to its dependency on $p_{\nu}(b)$, which in turn depends on the duration of the previous frame(s), as well as on the previous transmission outcomes, therefore being time-variant and strictly dependent on the traffic generation statistics. In the following, for simplicity, we assume that the \ac{bs} only knows the number of active users and performs the time resource scheduling conditioned on this partial information.

From a practical perspective, since $M_1$ is a design parameter, it should be time-invariant and should not depend on the user activations statistics.
Hence, by defining $\bar{P}(\nu(t)) = \nu(t) + \sigma_w^2$ and choosing $\epsilon_b = \bar{P}(b) +\frac{1}{2}$, $\forall b$ we approximate \eqref{eq:perr} as 
\begin{equation}\label{eq:perrapprox}
    p_{\rm e}(b) \approx 2{\rm Q}\left(\frac{\sqrt{M_1}}{2(b+\sigma_w^2)}\right).
\end{equation}
As ${\rm Q}(\cdot)$ is a monotonically decreasing function, the  maximum estimation error probability is achieved if all users are active in the frame $t$. 
Then, we derive $M_1$ in this worst-case scenario, which provides a target error probability $\tilde{p}_{\rm e} = p_{\rm e}(K)$, as
\begin{equation}\label{perr_constraint}
    M_1 = \left[2(K + \sigma_w^2){\rm Q}^{-1}\left(\frac{\tilde{p}_{\rm e}}{2}\right)\right]^2.
\end{equation}

\section{Frame-Efficiency-Based Scheduling}\label{sec:frameeff}

In this section, we propose a time-resource scheduling conditioned on the number of active users estimated in the \ac{pia} sub-frame.
Let $l \in \{1,\ldots, L_2(t)\}$ be the slot index within frame $t$ (in the \ac{dt} sub-frame). We define the success indicator function at slot $l$ as $c_l = 1$, if a successful transmission occurs at slot $l$ and $c_l = 0$ otherwise. Then, the \textit{conditional frame efficiency} is defined as the ratio between the number of successes in frame $t$ and the length of the \ac{dt} sub-frame, i.e., 
\begin{equation}\label{frameeff}
    \eta(t) = \frac{1}{L_2(t)}\sum_{l=1}^{L_2(t)}\mathbb{E}[c_l|\nu(t)].
\end{equation}
The adaptive maximization of this metric provides the proper balance between the \ac{dt} sub-frame length and the successful transmission probability.

In frame $t$, immediately after the end of the \ac{pia}, the \ac{bs} solves the following optimization problem:
\begin{subequations}\label{feopt}
	\label{maxprob}
	\begin{equation}
\max_{\substack{\bm{q}(t)}}\eta(t),
	\end{equation}
	\begin{equation}
{\rm     s.t.\;}		\;\; q_{k}(t) \in \{1,\ldots, L_2(t)\}.
	\end{equation}
\end{subequations}
Note that, without making any assumption on the activation probability distribution, the problem is not solvable in closed form; therefore we focus on the case of i.i.d. activation probabilities are equally distributed, without considering the correlations caused by the retransmission attempts and the packets generated during the \ac{dt} sub-frame. 

First, we observe that, since activations of users are i.i.d., we only have to determine how many users are assigned to each slot, as any specific assignment satisfying this constraint will yield the same collision probabilities and thus the same expected frame efficiency.

To minimize the number of users assigned to the same slot, given a length $L_2(t)$, we assign to slot $l$ the following number of users
\begin{equation}\label{udef}
	u_{l}(t) = 
	\begin{dcases}
		\left\lceil\frac{K}{L_2(t)}\right\rceil \quad {\rm if} \quad l \leq K\bmod{L_2(t)},\\
		\left\lfloor\frac{K}{L_2(t)}\right\rfloor \quad {\rm if} \quad l > K\bmod{L_2(t)},
	\end{dcases}
\end{equation}
where we may schedule one more user in the first $\left\lfloor\frac{K}{L_2(t)}\right\rfloor$ slots to minimize the transmission delay.

Note that the slot success random variable $c_l$ can be rewritten as a function of $u_l(t)$, as it only depends on the number of users scheduled in slot $l$. Now, the optimization problem \eqref{feopt} is reduced to the optimization of the second sub-frame length, $L_2(t)$, i.e., from \eqref{frameeff}, we have
\begin{subequations}
	\label{iidproblem}
	\begin{equation}
		L_2^*(t) = \argmax_{\substack{L_2(t)}} \frac{1}{L_2(t)}\sum_{l=1}^{L_2(t)}\mathbb{E}[c_{l}|\nu(t), u_l(t)],
	\end{equation}
 	\begin{equation}
		\mbox{s.t. } L_2(t) \in \mathbb{K}/\{0\}.\label{constraintISA}
	\end{equation}
\end{subequations}

Now, given $\nu(t)$, the probability of user $k$ being the one and only active user assigned to slot $l$ is derived by considering all cases of active users, where user $k$ is active and all other users assigned to slot $l$ are, instead, inactive. The number of favorable cases is given by all the possibilities to put $\nu(t)-1$ objects on a chessboard with $K-u_l(t)$ places, i.e., all the combinations of $\nu(t)-1$ elements taken from $K-u_l(t)$. For each combination, there are $K-u_l(t)$ possibilities to put the active user in slot $l$. 
Therefore, the collision probability in slot $l$ is $1 - \mathbb{E}[c_{l}|\nu(t), u_l(t)]$, where 
\begin{equation}\label{Ec_l}
	\mathbb{E}[c_{l}|\nu(t), u_l(t)] = \frac{u_l(t)\binom{K-u_l(t)}{\nu(t)-1}}{\binom{K}{\nu(t)}},
\end{equation}
is the probability of having a successful transmission in slot $l$.
Note that, in \eqref{Ec_l}, the numerator counts the number of combinations giving exactly one active user assigned to slot $l$, while the denominator counts the total number of possible combinations of active users.

Note that \eqref{iidproblem} is a \ac{minlp} problem, and is not solvable by continuous relaxation of $L_2(t)$, as the rounding functions are not differentiable. However, it is possible to find the optimal frame length $L_2^{*}(t)$ with complexity $O(\log K)$, using a binary search algorithm. In any case, $L_2^*(t)$ depends only on $\nu(t)$, thus can be computed offline and then stored in a table.

\section{Numerical Results}\label{sec:numericalresults}

In this section, we present the numerical results, comparing our \ac{pima} approaches with the \ac{tdma} and \ac{saloha} schedulers. 

We assume that the traffic generation, also denoted as {\em packet arrival process}, at each user follows a Poisson distribution with parameter $\lambda$. We consider $K=20$ users. The well-known properties of the Poisson processes provide a total arrival rate of $\Lambda = K\lambda$.
For a performance comparison, we consider a) the standard \ac{tdma}, which provides frames of fixed duration of $K$ slots, with one user assigned per slot, deterministically, and b) the \ac{saloha} protocol. In basic \ac{saloha}, in every time interval $l$, users transmit their packets immediately upon generation unless they are {\em backlogged} after a collision, in which case they transmit with a backoff probability. Instead, we consider Rivest's stabilized \ac{saloha} \cite[Chapter~4]{bertsekas21}, wherein all users generating packets in slot $l$ are backlogged with the same backoff probability. The backoff probability is computed for each user through a pseudo-Bayesian algorithm based on an estimate of the number of backlogged nodes $G(l)$ as 
\begin{equation}
    \alpha(l) = \min\left(1, \frac{1}{G(l)}\right),
 \end{equation}
where 
\begin{equation}
    G(l) = \begin{cases}
        G(l-1) + N\theta + (e-2)^{-1}\quad \text{if } c_{l-1} = 0,\\
        \max(N\theta, G(l-1)+ N\theta-1) \quad \text{if } c_{l-1} = 1,
    \end{cases}
\end{equation}
is the estimated number of users backlogged (with $G(0) = 0$) and $\theta$ is the probability packet generation in slot $l$.
Note that, despite being conventional multiple access solutions, \ac{saloha} and \ac{tdma} remain the ideal solutions for \ac{oma} in the time domain in case of low and high traffic conditions, respectively.
In the following, the comparison is made in terms of packet loss probability and average packet latency.

\subsection{PIA Parameters and Results}

Although the analysis of Section~\ref{sec:frameeff} assumes a perfect estimation of the number of active users, results shown in this section are obtained using the estimated $\hat{\nu}(t)$, thus including the effects of an estimation error.

In the \ac{pia} sub-frame, $M_1$ has to be large enough to make the approximation \eqref{perr_constraint} valid and ensure low estimation error probability \eqref{eq:perrapprox}. 
During the \ac{pia} sub-frame we assume that \ac{ai} signals transmitted by the active users have unitary power, while the noise power is $\sigma_w^2 = -10$~dB.
Furthermore, we adopt the third numerology of the \ac{nr} specification, which provides \ac{dt} time slots of $0.125$~ms \cite{3GPP38211}, and assume a total system bandwidth $W = 100$~MHz. 
Moreover, we neglect the duration of downlink beacons.
In the following, either $L_1 = 17$~$\mu$s or 44~$\mu$s, such that the target error probability \eqref{maxprob} is either $\tilde{p}_{\rm e}= 0.1$ or $\tilde{p}_{\rm e}=0.3$, respectively.

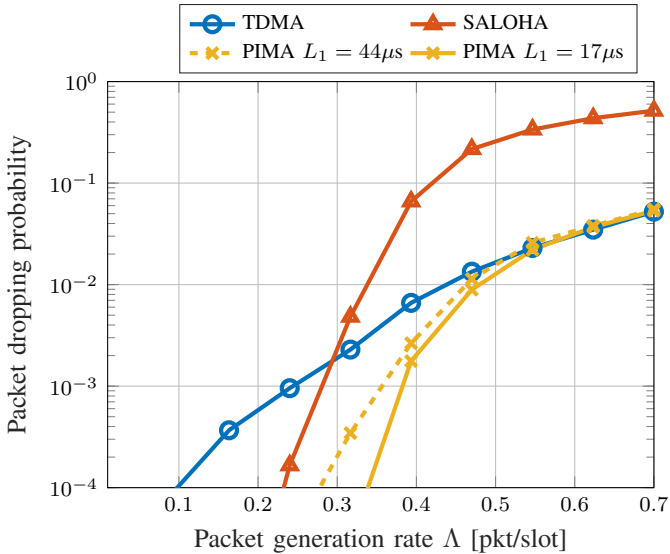
\begin{figure}
    \centering
    \setlength\fwidth{0.82\columnwidth}
    \setlength\fheight{0.61\columnwidth}
    \definecolor{mycolor1}{rgb}{0.00000,0.44700,0.74100}%
\definecolor{mycolor2}{rgb}{0.85000,0.32500,0.09800}%
\definecolor{mycolor3}{rgb}{0.92900,0.69400,0.12500}%
\pgfplotsset{every tick label/.append style={font=\footnotesize}}

\begin{tikzpicture}

\begin{axis}[%
    width=\fwidth,
    height=\fheight,
    at={(0\fwidth,0\fheight)},
    scale only axis,
    ylabel style={font=\normalsize},
    xlabel style={font=\normalsize},
    xmin=0.01,
    xmax=0.7,
    xlabel style={font=\color{white!15!black}},
    xlabel={Packet generation rate $\Lambda$ [pkt/slot]},
    ymode=log,
    ymin=1e-4,
    ymax=1,
    yminorticks=true,
    ylabel style={font=\color{white!15!black}},
    ylabel={Packet dropping probability},
    axis background/.style={fill=white},
    xmajorgrids,
    ymajorgrids,
    legend style={at={(0.97,1.02)}, anchor=south east, legend columns=2, legend cell align=left, align=left, font=\footnotesize, draw=white!15!black}
]

\addplot [color=mycolor1, ultra thick, mark size=3.0pt, mark=o, mark options={solid, mycolor1}]
  table[row sep=crcr]{%
0.01	0.000003\\
0.0866666666666667	0.00008\\
0.163333333333333	0.000367489434678753\\
0.24	0.00095475300954753\\
0.316666666666667	0.00229120241047048\\
0.393333333333333	0.0066128218071681\\
0.47	0.0134070899225134\\
0.546666666666667	0.0229573259539761\\
0.623333333333333	0.0347966337171236\\
0.7	0.0523457492180836\\
};
\addlegendentry{TDMA}

\addplot [color=mycolor2, ultra thick, mark size=3.0pt, mark=triangle, mark options={solid, mycolor2}]
  table[row sep=crcr]{%
0.01	0\\
0.0866666666666667	0\\
0.163333333333333	0.000001\\
0.24	0.000166037109293927\\
0.316666666666667	0.00483273708654993\\
0.393333333333333	0.0659986875977992\\
0.47	0.216435869289876\\
0.546666666666667	0.337265610777353\\
0.623333333333333	0.435818416976703\\
0.7	0.516660506638615\\
};
\addlegendentry{SALOHA}

\addplot [color=mycolor3, ultra thick, dashed, mark size=3.0pt, mark=x, mark options={solid, mycolor3}]
  table[row sep=crcr]{%
0.01	0\\
0.0866666666666667	0\\
0.163333333333333	0\\
0.24	0.00003\\
0.316666666666667	0.000345195506182138\\
0.393333333333333	0.00265010979026274\\
0.47	0.01137874981528\\
0.546666666666667	0.0258529658085424\\
0.623333333333333	0.0379892371812273\\
0.7	0.055286240279772\\
};
\addlegendentry{PIMA $L_1 =44\mu$s}

\addplot [color=mycolor3, ultra thick,  mark size=3.0pt, mark=x, mark options={solid, mycolor3}]
  table[row sep=crcr]{%
0.01	0\\
0.0866666666666667	0\\
0.163333333333333	0\\
0.24	0\\
0.316666666666667	3.13814096529216e-05\\
0.393333333333333	0.00176669527030438\\
0.47	0.00890878000379995\\
0.546666666666667	0.0220462406699436\\
0.623333333333333	0.0369181636726547\\
0.7	0.0534112395331315\\
};
\addlegendentry{PIMA $L_1 =17\mu$s}

\end{axis}
\end{tikzpicture}%
    \caption{Average packet dropping probability versus the total packet generation rate, for $K = 20$ and $\bar{B}=3$.}
    \label{fig:ploss}
\end{figure}

In Fig.~\ref{fig:ploss}, we show the empirical packet dropping probability as a function of the packet generation rate $\Lambda$, for $K=20$ users with buffer length $\bar{B}=3$.
Such metric measures the probability of packet replacement in the buffers when fresher packets are generated at users.
In low-traffic conditions, very few packets are dropped for both \ac{saloha} and \ac{pima}. A higher number of packets are dropped with \ac{tdma}, as the constantly adopted maximum frame length leads to time waste. 
In high-traffic conditions, instead, the dropping probability provided by \ac{saloha} keeps growing, performing even worse than \ac{tdma}, as the backlogging system prevents most users from transmitting to avoid collisions.
The advantages of both conventional schemes are merged in \ac{pima}, which is able to adapt to traffic conditions to provide fewer packet drops regardless of the intensity of packet generation. In the worst-case scenario, \ac{pima} matches \ac{tdma}, with a slight gap due to the overhead of \ac{pia}. 

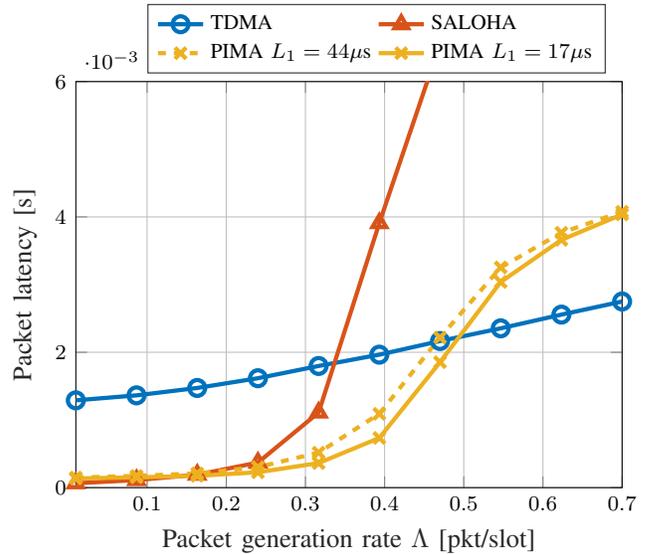
\begin{figure}
    \centering
       \setlength\fwidth{0.82\columnwidth}
    \setlength\fheight{0.61\columnwidth}
    \definecolor{mycolor1}{rgb}{0.00000,0.44700,0.74100}%
\definecolor{mycolor2}{rgb}{0.85000,0.32500,0.09800}%
\definecolor{mycolor3}{rgb}{0.92900,0.69400,0.12500}%
\pgfplotsset{every tick label/.append style={font=\footnotesize}}

\begin{tikzpicture}

\begin{axis}[%
    width=\fwidth,
    height=\fheight,
    at={(0\fwidth,0\fheight)},
    scale only axis,
    ylabel style={font=\normalsize},
    xlabel style={font=\normalsize},
    xmin=0.01,
    xmax=0.7,
    xlabel style={font=\color{white!15!black}},
    xlabel={Packet generation rate $\Lambda$ [pkt/slot]},
    ymin=0,
    ymax=0.006,
    yminorticks=true,
    ylabel style={font=\color{white!15!black}},
    ylabel={Packet latency [s]},
    axis background/.style={fill=white},
    xmajorgrids,
    ymajorgrids,
    legend style={at={(0.97,1.02)}, anchor=south east, legend columns=2, legend cell align=left, align=left, font=\footnotesize, draw=white!15!black}
]

\addplot [color=mycolor1, ultra thick, mark size=3.0pt, mark=o, mark options={solid, mycolor1}]
  table[row sep=crcr]{%
0.01	0.00128986055088915\\
0.0866666666666667	0.00136367131998872\\
0.163333333333333	0.00147281340038201\\
0.24	0.00161662982953164\\
0.316666666666667	0.00179773534056201\\
0.393333333333333	0.00196475706825112\\
0.47	0.0021660177641874\\
0.546666666666667	0.00235387481457237\\
0.623333333333333	0.00255662204795279\\
0.7	0.00275037339767452\\
};
\addlegendentry{TDMA}

\addplot [color=mycolor2, ultra thick, mark size=3.0pt, mark=triangle, mark options={solid, mycolor2}]
  table[row sep=crcr]{%
0.01	6.4810898456182e-05\\
0.0866666666666667	0.000111514137152902\\
0.163333333333333	0.00019129613195551\\
0.24	0.000370442361615462\\
0.316666666666667	0.00111025336538635\\
0.393333333333333	0.00390903200050316\\
0.47	0.00655735553445788\\
0.546666666666667	0.00749815104276447\\
0.623333333333333	0.00764534754777619\\
0.7	0.00749429891036252\\
};
\addlegendentry{SALOHA}

\addplot [color=mycolor3, ultra thick, dashed, mark size=3.0pt, mark=x, mark options={solid, mycolor3}]
  table[row sep=crcr]{%
0.01	0.000148140929409667\\
0.0866666666666667	0.00017313117727551\\
0.163333333333333	0.000210717695110196\\
0.24	0.000303196971028858\\
0.316666666666667	0.000517238824416354\\
0.393333333333333	0.00109065522229288\\
0.47	0.00221851206222394\\
0.546666666666667	0.00325369726841627\\
0.623333333333333	0.0037666217580183\\
0.7	0.00407651908713029\\
};
\addlegendentry{PIMA $L_1 =44\mu$s}

\addplot [color=mycolor3, ultra thick,  mark size=3.0pt, mark=x, mark options={solid, mycolor3}]
  table[row sep=crcr]{%
0.01	0.000135199128942133\\
0.0866666666666667	0.000152263503098448\\
0.163333333333333	0.000174575366777553\\
0.24	0.000227156759068874\\
0.316666666666667	0.00036292761033583\\
0.393333333333333	0.000734538196762533\\
0.47	0.00185392449461982\\
0.546666666666667	0.00303871877217582\\
0.623333333333333	0.00366116106157138\\
0.7	0.00403662165583288\\
};
\addlegendentry{PIMA $L_1 =17\mu$s}

\end{axis}
\end{tikzpicture}%
    \caption{Average packet latency versus the total packet generation rate, for $K = 20$ and $\bar{B}=3$.}
    \label{fig:latency}
\end{figure}

Finally, Fig.~\ref{fig:latency} shows the impact of the packet generation rate on the average latency. As expected, the \ac{tdma}  and \ac{saloha} protocols perform well at low and high packet generation rates, respectively. In fact, at low traffic intensity \ac{tdma} suffers from a long waiting due to deterministic slot allocation. \ac{saloha} instead has a very low latency at low traffic, as packets are immediately transmitted, while more collisions occur as the probability of packet generation increases, increasing the latency. Again, \ac{pima} merges the two advantages to achieve great results in all traffic conditions. However, a slight performance degradation is observed at high traffic, due to both the \ac{pia} overhead and the additional delay introduced when the packet is delayed to the next frame upon generation.

Finally, note that as $L_1$ decreases, packet loss and latency are reduced. However, the short \ac{pia} duration of the subframe could imply a lower $M_1$, leading to erroneous estimation of $\nu(t)$, and thus making the \ac{bs} perform inefficient scheduling.

\section{Conclusions}\label{sec:conclusions}

For addressing the challenging requirements of the emerging \ac{5g}/\ac{6g} \ac{iot} use cases, we have proposed a new semi-\ac{gf} coordinated multiple access scheme, the \ac{pima} protocol, based on the knowledge of the number of users that have packets to transmit. To this end, \ac{pima} organizes time into frames, and each frame includes a preliminary phase (the \ac{pia} sub-frame), wherein active users transmit a compute-over-the-air signal that enables the \ac{bs} to estimate the number of active users. Then, we analyzed the protocol, indicating the policies to be used to minimize packet losses and latency. 
From the analysis and the numerical results obtained in a generic \ac{mtc} scenario, we conclude that \ac{pima} with proper scheduling significantly reduces packet loss and latency with respect to \ac{tdma} and \ac{saloha}, particularly when dealing with sporadic activations. 

\balance

\bibliographystyle{IEEEtran}
\bibliography{IEEEabrv, Bibliography}

\end{document}